\documentstyle[twoside,fleqn,espcrc2]{article}

\newcommand{\skipover}[1]{}
\newcommand{\be}{\begin{equation}}
\newcommand{\ee}{\end{equation}}
\newcommand{\bearray}{\begin{eqnarray}}
\newcommand{\eearray}{\end{eqnarray}}
\newcommand{\nl}{\nonumber \\}
\newcommand{\pl}{{\rm pl}}
\newcommand{\rt}{{\rm rt}}
\newcommand{\pg}{{\rm pg}}
\newcommand{\ls}[1]{\mbox{$\frac{1}{3}$\,Re\,Tr}(1-#1)}
\newcommand{\order}{{\cal O}}

\title{QCD on Coarse Lattices${}^*$}

\author{M.~Alford, W.~Dimm and G.P.~Lepage \\
{\sl Floyd R. Newman Laboratory of Nuclear Studies,
       Cornell University, Ithaca, NY 14853} \\[0.05in]
G.~Hockney and P.B.~Mackenzie \\
{\sl Theoretical Physics Department, Fermilab, Batavia, IL 60510}\\[0.1in]
We show that the perturbatively-improved gluon action for QCD, once it
is tadpole-improved, gives accurate results even with lattice spacings
as large as 0.4~fm. {\em No\/} tuning of the couplings is required.
Using this action and lattice spacing,
we obtain a static potential that is rotationally
invariant to within a few percent,
the spin-averaged charmonium spectrum accurate to within 30--40~MeV, and
scaling  to within 5--10\%. We demonstrate that simulations on
coarse lattices
are several orders of magnitude less costly than simulations using
current methods.\\[0.1in]
${}^*$ Talk presented by G.P.~Lepage at {\em Lattice '94} (Bielefeld,
Germany, September 1994)}

\begin{document}
\maketitle

Recent studies indicate that weak-coupling perturbation theory, when properly
renormalized and tadpole improved, gives an accurate
account of lattice-QCD dynamics even at distances as large as
$1/2$~fm~\cite{lm}. This suggests that accurate
simulations can be carried out on very coarse lattices, perhaps with lattice
spacing~$a\approx1/2$~fm, by using (asymptotically free) perturbation theory to
correct for dynamics at distance scales smaller than~$a$.
In this paper we present new evidence that this is
indeed the case for gluon dynamics. We show accurate results
obtained with spacing~$a\approx 0.4$~fm. This is 4--8~times larger than
usual~$a$'s, and leads to a spectacular reduction in simulation costs, which
typically vary as $1/a^6$.

The standard discretizations of the quark and gluon actions are not
sufficiently accurate for simulations on such coarse lattices; the leading
finite-$a$ errors must be removed. Such errors can be
systematically removed from on-shell quantities, like scattering amplitudes and
masses, by adding nonrenormalizable terms to the lattice actions~\cite{sym}.
Since these terms correct primarily the short-distance behavior of the theory,
their coefficients can be computed using perturbation theory once~$a$ is
sufficiently small. Early tests of this technique for lattice QCD showed
little improvement, but these used naive lattice perturbation theory and so
seriously underestimated the coefficients.  The use of renormalized
perturbation
theory and especially tadpole improvement results in significantly larger
coefficients and, as we will show, greatly reduced finite-$a$ errors.
Note that, unlike some improvement schemes, perturbative improvement of this
sort does {\em not} introduce new parameters to be tuned numerically; here the
additional couplings in the action are computed, before the
simulation, using tadpole-improved perturbation theory.

Perturbatively improved actions have already proven very successful in
simulations of heavy-quark mesons like the~$\Upsilon$. In the
nonrelativistic NRQCD quark action, both relativistic effects and
finite-$a$ corrections are introduced through nonrenormalizable corrections to
the basic action. The
detailed simulation results presented in~\cite{nrqcd} agree well with
experiment, and many depend
crucially on these corrections.
Tadpole improvement was essential to
this success; test runs without tadpole improvement underestimated relativistic
effects by as much as a factor of two.

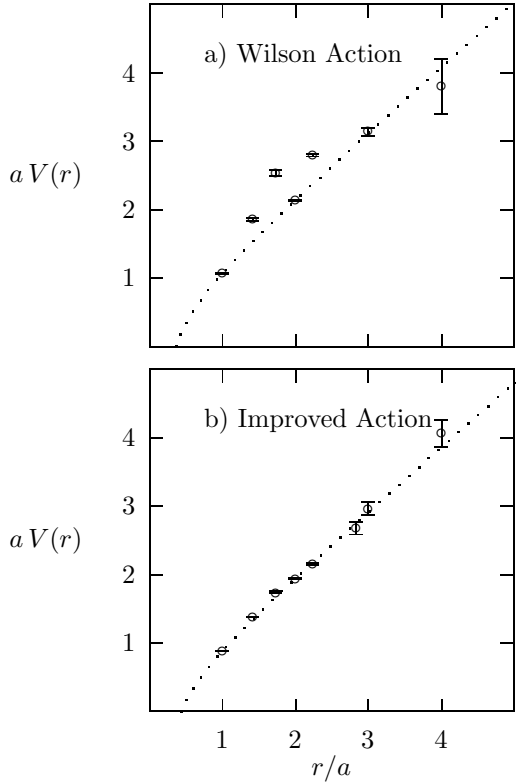
\begin{figure}[t]

\mbox{}\vspace{-0.3in}

\setlength{\unitlength}{0.240900pt}
\ifx\plotpoint\undefined\newsavebox{\plotpoint}\fi
\sbox{\plotpoint}{\rule[-0.175pt]{0.350pt}{0.350pt}}%
\begin{picture}(900,809)(0,0)
\tenrm
\sbox{\plotpoint}{\rule[-0.175pt]{0.350pt}{0.350pt}}%
\put(264,158){\rule[-0.175pt]{137.795pt}{0.350pt}}
\put(264,158){\rule[-0.175pt]{0.350pt}{129.604pt}}
\put(264,266){\rule[-0.175pt]{4.818pt}{0.350pt}}
\put(242,266){\makebox(0,0)[r]{$1$}}
\put(816,266){\rule[-0.175pt]{4.818pt}{0.350pt}}
\put(264,373){\rule[-0.175pt]{4.818pt}{0.350pt}}
\put(242,373){\makebox(0,0)[r]{$2$}}
\put(816,373){\rule[-0.175pt]{4.818pt}{0.350pt}}
\put(264,481){\rule[-0.175pt]{4.818pt}{0.350pt}}
\put(242,481){\makebox(0,0)[r]{$3$}}
\put(816,481){\rule[-0.175pt]{4.818pt}{0.350pt}}
\put(264,588){\rule[-0.175pt]{4.818pt}{0.350pt}}
\put(242,588){\makebox(0,0)[r]{$4$}}
\put(816,588){\rule[-0.175pt]{4.818pt}{0.350pt}}
\put(378,158){\rule[-0.175pt]{0.350pt}{4.818pt}}
\put(378,676){\rule[-0.175pt]{0.350pt}{4.818pt}}
\put(493,158){\rule[-0.175pt]{0.350pt}{4.818pt}}
\put(493,676){\rule[-0.175pt]{0.350pt}{4.818pt}}
\put(607,158){\rule[-0.175pt]{0.350pt}{4.818pt}}
\put(607,676){\rule[-0.175pt]{0.350pt}{4.818pt}}
\put(722,158){\rule[-0.175pt]{0.350pt}{4.818pt}}
\put(722,676){\rule[-0.175pt]{0.350pt}{4.818pt}}
\put(264,158){\rule[-0.175pt]{137.795pt}{0.350pt}}
\put(836,158){\rule[-0.175pt]{0.350pt}{129.604pt}}
\put(264,696){\rule[-0.175pt]{137.795pt}{0.350pt}}
\put(45,427){\makebox(0,0)[l]{\shortstack{$a\,V(r)$}}}
\put(350,615){\makebox(0,0)[l]{a) Wilson Action}}
\put(264,158){\rule[-0.175pt]{0.350pt}{129.604pt}}
\put(378,274){\circle{12}}
\put(426,358){\circle{12}}
\put(462,431){\circle{12}}
\put(493,388){\circle{12}}
\put(520,459){\circle{12}}
\put(607,496){\circle{12}}
\put(722,567){\circle{12}}
\put(378,273){\usebox{\plotpoint}}
\put(368,273){\rule[-0.175pt]{4.818pt}{0.350pt}}
\put(368,274){\rule[-0.175pt]{4.818pt}{0.350pt}}
\put(426,355){\rule[-0.175pt]{0.350pt}{1.445pt}}
\put(416,355){\rule[-0.175pt]{4.818pt}{0.350pt}}
\put(416,361){\rule[-0.175pt]{4.818pt}{0.350pt}}
\put(462,427){\rule[-0.175pt]{0.350pt}{2.168pt}}
\put(452,427){\rule[-0.175pt]{4.818pt}{0.350pt}}
\put(452,436){\rule[-0.175pt]{4.818pt}{0.350pt}}
\put(493,387){\rule[-0.175pt]{0.350pt}{0.482pt}}
\put(483,387){\rule[-0.175pt]{4.818pt}{0.350pt}}
\put(483,389){\rule[-0.175pt]{4.818pt}{0.350pt}}
\put(520,457){\rule[-0.175pt]{0.350pt}{0.964pt}}
\put(510,457){\rule[-0.175pt]{4.818pt}{0.350pt}}
\put(510,461){\rule[-0.175pt]{4.818pt}{0.350pt}}
\put(607,489){\rule[-0.175pt]{0.350pt}{3.132pt}}
\put(597,489){\rule[-0.175pt]{4.818pt}{0.350pt}}
\put(597,502){\rule[-0.175pt]{4.818pt}{0.350pt}}
\put(722,524){\rule[-0.175pt]{0.350pt}{20.717pt}}
\put(712,524){\rule[-0.175pt]{4.818pt}{0.350pt}}
\put(712,610){\rule[-0.175pt]{4.818pt}{0.350pt}}
\sbox{\plotpoint}{\rule[-0.250pt]{0.500pt}{0.500pt}}%
\put(305,158){\usebox{\plotpoint}}
\put(312,177){\usebox{\plotpoint}}
\put(321,195){\usebox{\plotpoint}}
\put(331,213){\usebox{\plotpoint}}
\put(344,230){\usebox{\plotpoint}}
\put(355,247){\usebox{\plotpoint}}
\put(369,263){\usebox{\plotpoint}}
\put(383,278){\usebox{\plotpoint}}
\put(397,294){\usebox{\plotpoint}}
\put(411,309){\usebox{\plotpoint}}
\put(426,324){\usebox{\plotpoint}}
\put(440,338){\usebox{\plotpoint}}
\put(455,352){\usebox{\plotpoint}}
\put(470,367){\usebox{\plotpoint}}
\put(485,381){\usebox{\plotpoint}}
\put(500,395){\usebox{\plotpoint}}
\put(515,410){\usebox{\plotpoint}}
\put(531,423){\usebox{\plotpoint}}
\put(546,438){\usebox{\plotpoint}}
\put(561,452){\usebox{\plotpoint}}
\put(576,465){\usebox{\plotpoint}}
\put(592,480){\usebox{\plotpoint}}
\put(607,494){\usebox{\plotpoint}}
\put(622,507){\usebox{\plotpoint}}
\put(638,521){\usebox{\plotpoint}}
\put(653,535){\usebox{\plotpoint}}
\put(668,549){\usebox{\plotpoint}}
\put(684,563){\usebox{\plotpoint}}
\put(699,577){\usebox{\plotpoint}}
\put(715,591){\usebox{\plotpoint}}
\put(730,604){\usebox{\plotpoint}}
\put(746,618){\usebox{\plotpoint}}
\put(761,632){\usebox{\plotpoint}}
\put(777,646){\usebox{\plotpoint}}
\put(792,660){\usebox{\plotpoint}}
\put(808,673){\usebox{\plotpoint}}
\put(823,688){\usebox{\plotpoint}}
\put(833,696){\usebox{\plotpoint}}
\end{picture}

 \vspace{-0.8in}

\setlength{\unitlength}{0.240900pt}
\ifx\plotpoint\undefined\newsavebox{\plotpoint}\fi
\sbox{\plotpoint}{\rule[-0.175pt]{0.350pt}{0.350pt}}%
\begin{picture}(900,809)(0,0)
\tenrm
\sbox{\plotpoint}{\rule[-0.175pt]{0.350pt}{0.350pt}}%
\put(264,158){\rule[-0.175pt]{137.795pt}{0.350pt}}
\put(264,158){\rule[-0.175pt]{0.350pt}{129.604pt}}
\put(264,266){\rule[-0.175pt]{4.818pt}{0.350pt}}
\put(242,266){\makebox(0,0)[r]{$1$}}
\put(816,266){\rule[-0.175pt]{4.818pt}{0.350pt}}
\put(264,373){\rule[-0.175pt]{4.818pt}{0.350pt}}
\put(242,373){\makebox(0,0)[r]{$2$}}
\put(816,373){\rule[-0.175pt]{4.818pt}{0.350pt}}
\put(264,481){\rule[-0.175pt]{4.818pt}{0.350pt}}
\put(242,481){\makebox(0,0)[r]{$3$}}
\put(816,481){\rule[-0.175pt]{4.818pt}{0.350pt}}
\put(264,588){\rule[-0.175pt]{4.818pt}{0.350pt}}
\put(242,588){\makebox(0,0)[r]{$4$}}
\put(816,588){\rule[-0.175pt]{4.818pt}{0.350pt}}
\put(378,158){\rule[-0.175pt]{0.350pt}{4.818pt}}
\put(378,113){\makebox(0,0){$1$}}
\put(378,676){\rule[-0.175pt]{0.350pt}{4.818pt}}
\put(493,158){\rule[-0.175pt]{0.350pt}{4.818pt}}
\put(493,113){\makebox(0,0){$2$}}
\put(493,676){\rule[-0.175pt]{0.350pt}{4.818pt}}
\put(607,158){\rule[-0.175pt]{0.350pt}{4.818pt}}
\put(607,113){\makebox(0,0){$3$}}
\put(607,676){\rule[-0.175pt]{0.350pt}{4.818pt}}
\put(722,158){\rule[-0.175pt]{0.350pt}{4.818pt}}
\put(722,113){\makebox(0,0){$4$}}
\put(722,676){\rule[-0.175pt]{0.350pt}{4.818pt}}
\put(264,158){\rule[-0.175pt]{137.795pt}{0.350pt}}
\put(836,158){\rule[-0.175pt]{0.350pt}{129.604pt}}
\put(264,696){\rule[-0.175pt]{137.795pt}{0.350pt}}
\put(45,427){\makebox(0,0)[l]{\shortstack{$a\,V(r)$}}}
\put(550,68){\makebox(0,0){$r/a$}}
\put(350,615){\makebox(0,0)[l]{b) Improved Action}}
\put(264,158){\rule[-0.175pt]{0.350pt}{129.604pt}}
\put(378,253){\circle{12}}
\put(426,307){\circle{12}}
\put(462,345){\circle{12}}
\put(493,367){\circle{12}}
\put(520,390){\circle{12}}
\put(588,446){\circle{12}}
\put(607,476){\circle{12}}
\put(722,595){\circle{12}}
\put(378,253){\usebox{\plotpoint}}
\put(368,253){\rule[-0.175pt]{4.818pt}{0.350pt}}
\put(368,253){\rule[-0.175pt]{4.818pt}{0.350pt}}
\put(426,307){\usebox{\plotpoint}}
\put(416,307){\rule[-0.175pt]{4.818pt}{0.350pt}}
\put(416,307){\rule[-0.175pt]{4.818pt}{0.350pt}}
\put(462,344){\rule[-0.175pt]{0.350pt}{0.723pt}}
\put(452,344){\rule[-0.175pt]{4.818pt}{0.350pt}}
\put(452,347){\rule[-0.175pt]{4.818pt}{0.350pt}}
\put(493,366){\rule[-0.175pt]{0.350pt}{0.482pt}}
\put(483,366){\rule[-0.175pt]{4.818pt}{0.350pt}}
\put(483,368){\rule[-0.175pt]{4.818pt}{0.350pt}}
\put(520,388){\rule[-0.175pt]{0.350pt}{0.723pt}}
\put(510,388){\rule[-0.175pt]{4.818pt}{0.350pt}}
\put(510,391){\rule[-0.175pt]{4.818pt}{0.350pt}}
\put(588,436){\rule[-0.175pt]{0.350pt}{4.818pt}}
\put(578,436){\rule[-0.175pt]{4.818pt}{0.350pt}}
\put(578,456){\rule[-0.175pt]{4.818pt}{0.350pt}}
\put(607,466){\rule[-0.175pt]{0.350pt}{5.059pt}}
\put(597,466){\rule[-0.175pt]{4.818pt}{0.350pt}}
\put(597,487){\rule[-0.175pt]{4.818pt}{0.350pt}}
\put(722,573){\rule[-0.175pt]{0.350pt}{10.359pt}}
\put(712,573){\rule[-0.175pt]{4.818pt}{0.350pt}}
\put(712,616){\rule[-0.175pt]{4.818pt}{0.350pt}}
\sbox{\plotpoint}{\rule[-0.250pt]{0.500pt}{0.500pt}}%
\put(313,158){\usebox{\plotpoint}}
\put(321,176){\usebox{\plotpoint}}
\put(332,194){\usebox{\plotpoint}}
\put(344,211){\usebox{\plotpoint}}
\put(356,228){\usebox{\plotpoint}}
\put(370,244){\usebox{\plotpoint}}
\put(383,259){\usebox{\plotpoint}}
\put(398,274){\usebox{\plotpoint}}
\put(412,289){\usebox{\plotpoint}}
\put(427,304){\usebox{\plotpoint}}
\put(441,318){\usebox{\plotpoint}}
\put(456,333){\usebox{\plotpoint}}
\put(471,347){\usebox{\plotpoint}}
\put(486,361){\usebox{\plotpoint}}
\put(501,375){\usebox{\plotpoint}}
\put(517,390){\usebox{\plotpoint}}
\put(532,403){\usebox{\plotpoint}}
\put(547,417){\usebox{\plotpoint}}
\put(563,431){\usebox{\plotpoint}}
\put(578,445){\usebox{\plotpoint}}
\put(593,459){\usebox{\plotpoint}}
\put(609,472){\usebox{\plotpoint}}
\put(624,487){\usebox{\plotpoint}}
\put(640,500){\usebox{\plotpoint}}
\put(655,514){\usebox{\plotpoint}}
\put(671,528){\usebox{\plotpoint}}
\put(687,541){\usebox{\plotpoint}}
\put(701,556){\usebox{\plotpoint}}
\put(717,569){\usebox{\plotpoint}}
\put(733,583){\usebox{\plotpoint}}
\put(748,597){\usebox{\plotpoint}}
\put(764,610){\usebox{\plotpoint}}
\put(779,624){\usebox{\plotpoint}}
\put(795,637){\usebox{\plotpoint}}
\put(810,652){\usebox{\plotpoint}}
\put(826,665){\usebox{\plotpoint}}
\put(836,674){\usebox{\plotpoint}}
\end{picture}

\vspace{-0.5in}

\caption{Static-quark potential computed on $6^4$ lattices with $a\approx
0.4$~fm using the $\beta=4.5$ Wilson action and the
improved action with $\beta_\pl = 6.8$,
$\beta_\rt = -0.562$, and $\beta_\pg= -0.0844$.}
\label{potl}
\end{figure}

The standard Wilson action for gluons has finite-$a$ errors of $\order(a^2)$,
which are plainly visible in the static-quark potential computed on
a lattice with $a\approx 0.4$~fm (see Fig.~\ref{potl}a). To reduce these errors
we use a corrected action~\cite{w} that includes two new terms in addition to
the usual plaquette term:
\bearray
S[U] &=& \beta_\pl \sum_\pl \ls{U_\pl} \nl
&+& \beta_\rt \sum_\rt \ls{U_\rt} \nl
&+& \beta_\pg \sum_\pg \ls{U_\pg} ,
\eearray
Here $U_\pl$ is the $1\times1$~plaquette operator, $U_\rt$ is the
$1\times2$~rectangle operator, and $U_\pg$ is the six-link
parallelogram operator
(path $\mu,\nu,\rho,-\mu,-\nu,-\rho$ where
$\mu,\nu,\rho$ are all different directions). In our work, we take the
plaquette
coupling $\beta_\pl$ as an input and compute the other two using
perturbation theory. By tadpole improving the one-loop results in~\cite{w} we
obtain:
\bearray
\beta_\rt &=& -\frac{\beta_\pl}{20\,u_0^2}\, \left( 1 + 0.4805\,\alpha_s
    \right) \\
\beta_\pg &=& -\frac{\beta_\pl}{u_0^2} \, 0.03325\,\alpha_s.
\eearray
As expected, tadpole improvement significantly improves the convergence of the
$\beta_\rt$~expansion. Following~\cite{lm}, we use the measured
expectation value of the plaquette to determine both the value of the mean
link~$u_0$ and the QCD coupling constant~$\alpha_s$ used
in these expressions:
\bearray
 u_0 &=& \left(\mbox{$\frac{1}{3}$\,Re\,Tr}\langle U_\pl \rangle\right)^{1/4},
\\
 \alpha_s &=& \frac{\ls{\langle U_\pl\rangle}}{3.06839}.
\eearray
The couplings $\beta_\rt$ and $\beta_\pg$ are determined self-consistently with
$u_0$ and~$\alpha_s$ for a given~$\beta_\pl$. As in NRQCD, there is no tuning
here of the couplings for the correction terms; tadpole-improved perturbation
theory determines them in terms of the single bare
coupling~$\beta_\pl$.\footnote{Using identies from~\cite{sym} we find that our
action is positive semidefinite at least for $\beta_\pl\ge6.8$.}

The static-quark potential computed using the improved gluon action is
shown in Fig.~\ref{potl}b. As in the Wilson case (Fig~\ref{potl}a),
the lattice spacing is about 0.4~fm.  The dashed line in these plots
is the standard infrared parameterization for the potential,
$V(r)=Kr-\pi/12r + c$, adjusted to fit the on-axis values of the
potential. Off-axis points deviate from the fit by as much as 35\% for
the Wilson theory, indicating a significant failure of rotation
invariance due to finite-$a$ errors. By contrast, the deviations are
only~2--4\% for the improved theory\,---\,negligible for most
low-energy applications. Tuning the couplings away from their one-loop
perturbative values results in little further improvement, suggesting
that $\order(a^4)$ effects are as important at this~$a$ as the
$\order(\alpha_s^2\,a^2)$ errors that have not been included.

To further check on  our improved theory, we examined the
spin-averaged spectrum of the $\psi$ family of mesons using NRQCD for the
$c$-quarks and our improved action at $\beta_\pl=$~6.8 and 7.4. Because we were
examining only the spin-averaged spectrum, we omitted all relativistic
corrections from the NRQCD action, but kept the corrections for
$\order(a,a^2)$~errors. The spectra, normalized to give the correct
$1P$-$1S$~splitting, are shown in Fig.~\ref{spect} together with experimental
results (dashed lines) and simulation results obtained using the Wilson action
for the gluons at smaller~$a$'s~\cite{nrqcd}. All agree to within 30--40~MeV.
Also the results scale well with the string tension~$\sigma$ in the
improved theory, but very poorly in the Wilson theory (see
Table~\ref{scaling}).
\begin{table}[t]
\begin{center}
\begin{tabular}{lcc|c}
& $\beta_\pl$ & $a$ & $(1P-1S)/\sqrt{\sigma}$ \\ \hline
Wilson & 4.5 & .41~fm & 1.46 (5) \\
& 5.7 & .17~fm & 0.92 (5) \\ \hline
Improved & 6.8 & .41~fm & 0.99 (5) \\
& 7.4 & .25~fm & 0.93 (5) \\
\end{tabular}
\end{center}
\caption{Ratio of the charmonium $1P-1S$~splitting to the square root of the
string tension.}
\label{scaling}
\end{table}
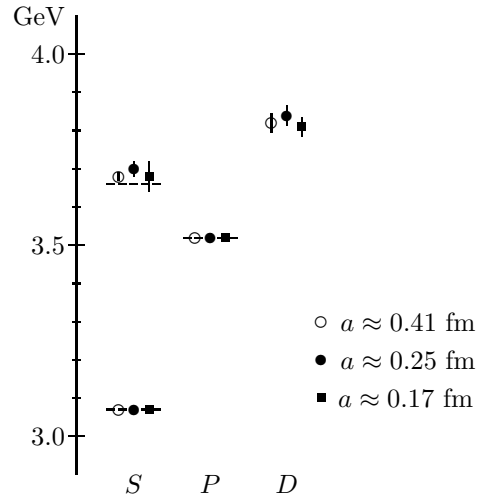
\begin{figure}
\begin{center}
\setlength{\unitlength}{.02in}
\begin{picture}(120,130)(0,280)
\put(79,330){\circle{3} \makebox(0,0)[l]{$a\approx0.41$~fm}}
\put(79,320){{\circle*{3}} \makebox(0,0)[l]{$a\approx0.25$~fm}}
\put(78,310){\rule[-\unitlength]{2\unitlength}{2\unitlength}
             \makebox(0,0)[l]{$\,a\approx0.17$~fm}}

\put(15,290){\line(0,1){120}}
\multiput(13,300)(0,50){3}{\line(1,0){4}}
\multiput(14,310)(0,10){9}{\line(1,0){2}}
\put(12,300){\makebox(0,0)[r]{3.0}}
\put(12,350){\makebox(0,0)[r]{3.5}}
\put(12,400){\makebox(0,0)[r]{4.0}}
\put(12,410){\makebox(0,0)[r]{GeV}}

\put(30,290){\makebox(0,0)[t]{$S$}}

\multiput(23,307)(3,0){5}{\line(1,0){2}}
\put(26,307){\circle{3}}
\put(30,307){\circle*{3}}
\put(33,306){\rule{2\unitlength}{2\unitlength}}

\multiput(23,366)(3,0){5}{\line(1,0){2}}
\put(26,368){\circle{3}}
\put(26,367){\line(0,1){2}}
\put(30,370){\circle*{3}}
\put(30,368){\line(0,1){4}}
\put(33,367){\rule{2\unitlength}{2\unitlength}}
\put(34,364){\line(0,1){8}}

\put(50,290){\makebox(0,0)[t]{$P$}}

\multiput(43,352)(3,0){5}{\line(1,0){2}}
\put(46,352){\circle{3}}
\put(50,352){\circle*{3}}
\put(50,351){\line(0,1){2}}
\put(53,351){\rule{2\unitlength}{2\unitlength}}
\put(54,351.5){\line(0,1){1}}

\put(70,290){\makebox(0,0)[t]{$D$}}

\put(66,382){\circle{3}}
\put(66,379.5){\line(0,1){5}}
\put(70,384){\circle*{3}}
\put(70,381.5){\line(0,1){5}}
\put(73,380){\rule{2\unitlength}{2\unitlength}}
\put(74,378.5){\line(0,1){5}}
\end{picture}

\vspace{-0.5in}

\end{center}
\caption{$S$, $P$, and $D$ states of charmonium computed on lattices with:
$a\approx0.41$~fm (improved action, $\beta_\pl=6.8$);
$a\approx0.25$~fm (improved action, $\beta_\pl=7.4$); and
$a\approx0.17$~fm (Wilson action, $\beta=5.7$). }
\label{spect}
\end{figure}

Since coarse lattices have far fewer sites and much
less critical-slowing-down, the cost to produce a statistically independent
configuration should be much less on a coarser lattice. To examine this issue,
we compared our results with those in~\cite{bs} for the potential computed
using
the Wilson action at
$\beta=6$ ($32^4$ lattice with $a\approx0.1$~fm). We rescaled the coordinates
and potential from this other study to put them in the same units as our
$\beta_\pl=6.8$ results, and examined the potentials at comparable distances.
The
results from both simulations are listed in Table~\ref{bspotl}. In both cases
the
potential  is obtained from the time dependence of loop-like correlation
functions for  times equal to or larger than  some~$T_{\rm min}$. Our results
required
$1.3\times10^7$ site updates, while the analysis on the fine lattice required
$6.4\times10^9$ site updates. Since statistical errors (for $T_{\rm
min}=0.4$~fm)
are about 20~times smaller for the coarse lattice, we estimate that comparable
errors with the fine lattice would require 197,000~times more site updates
than we used on the coarse lattice.
\begin{table}
\begin{center}
\begin{tabular}{c|ll|l}
 & \multicolumn{2}{c|}{$a=a_c\approx 0.41$} &
   \multicolumn{1}{c}{$a\approx0.10$} \\
 $r/a_c$ & $T_{\rm min}=.4$ & $T_{\rm min}=.8$ & $T_{\rm min}=.3$ \\ \hline
 1 & 0.871 ( 0) & 0.887 ( 1) & \\
 0.962 &&& 0.839 (11) \\
 $\sqrt{2}$ & 1.373 ( 1) & 1.384 ( 2) & \\
 1.361 &&& 1.316 (25) \\
 $\sqrt{3}$ & 1.718 ( 2) & 1.742 (10) & \\
 1.667 &&& 1.581 (54) \\
 2 & 1.897 ( 2) & 1.941 (10) & \\
 1.924 &&& 1.842 (42) \\
 \end{tabular}
 \caption{Comparison of the static-quark potential $a\,V(r)$  as computed on
a coarse lattice (improved action, $\beta_\pl=6.8$,
$a\approx 0.41$~fm), and on a fine lattice (Wilson action, $\beta=6$,
$a\approx 0.10$~fm).}
\label{bspotl}
\end{center}
 \end{table}

Comparisons of the sort just discussed are necessarily crude, but the
implications are clear. The shift to coarse lattices and perturbatively
improved
actions, with tadpole-improved couplings, makes possible the calculation of the
QCD hadron spectrum to, say, 1\% accuracy with orders of magnitude less
computing power than is possible with current methods.

We are very grateful to A.~Pierce, IBM, and Cornell's Center for
Theory and Simulation
for help in using the IBM SP-2 supercomputer for parts of
this work. This work was supported by the DOE and NSF.


\begin{thebibliography}{99}
\bibitem{lm} G.P. Lepage and P.B. Mackenzie, {\em Phys. Rev.\/} {\bf D48},
2250 (1993).

\bibitem{sym} M.~L\"uscher and P.~Weisz, {\em Comm. Math. Phys.\/} {\bf 97},
59~(1985); K.\ Symanzik, {\em Nucl.\ Phys.\/} {\bf B226}, 187(1983).

\bibitem{nrqcd} C.~Davies et al, to be published in
{\em  Phys.\ Rev.\/} {\bf D} (1994); and talks by C.~Davies and
J.~Sloan at this meeting.

\bibitem{w} See M.~L\"uscher and P.~Weisz, {\em Phys. Lett.\/} {\bf 158B},
250(1985), and references therein.

\bibitem{bs} G.S.~Bali and K.~Schilling, {\em  Phys.\ Rev.\/} {\bf D46},
2636(1992).

\end{thebibliography}
\end{document}